\documentclass[conference]{IEEEtran}
\IEEEoverridecommandlockouts
\usepackage{cite}
\usepackage{amsmath,amssymb,amsfonts}
\usepackage{algorithmic}
\usepackage{graphicx}
\usepackage{textcomp}
\usepackage{xcolor}
\usepackage[draft]{hyperref}
\usepackage{booktabs} 
\usepackage{pgfplots} 
\usepackage{multirow}

\pgfplotsset{compat=1.18}
\def\BibTeX{{\rm B\kern-.05em{\sc i\kern-.025em b}\kern-.08em
    T\kern-.1667em\lower.7ex\hbox{E}\kern-.125emX}}

\begin{document}

\title{CircuChain: Disentangling Competence and Compliance in LLM Circuit Analysis}

\author{\IEEEauthorblockN{Mayank Ravishankara}
\IEEEauthorblockA{\textit{Independent Researcher} \\
San Francisco, CA, USA \\
mravisha@alumni.cmu.edu (alt: mayankgowda@gmail.com)}
}

\maketitle

\begin{abstract}
As Large Language Models (LLMs) advance toward expert-level performance in engineering domains, reliable reasoning under user-specified constraints becomes critical. In circuit analysis, for example, a numerically correct solution is insufficient if it violates established methodological conventions such as mesh directionality or polarity assignments—errors that can propagate in safety-critical systems. Yet it remains unclear whether frontier models truly apply first-principles reasoning or rely on entrenched training priors that conflict with explicit instructions.

We introduce \textbf{CircuChain}, a diagnostic benchmark designed to disentangle \emph{instruction compliance} from \emph{physical reasoning competence} in electrical circuit analysis. CircuChain consists of counterbalanced Control/Trap problem pairs across five canonical circuit topologies, augmented with systematic variations in sign conventions, current orientations, and polarity definitions. A multi-stage verification pipeline—combining symbolic solvers, SPICE simulation, and an LLM-based error taxonomy—enables fine-grained attribution of failures to convention errors, physics errors, arithmetic mistakes, or hallucinations.

Across 100 tasks per model, we observe a consistent \textbf{Compliance–Competence Divergence}. The strongest model evaluated exhibits near-perfect physical reasoning but a high rate of convention violations when Trap conditions deliberately invert natural sign patterns. Conversely, weaker models display lower physical fidelity yet superior adherence to explicit instructions. These results suggest that increased model capability does not guarantee improved constraint alignment and highlight the need for new evaluation frameworks that stress instruction-following under mathematically rigid domains.

CircuChain provides one such framework and offers actionable insights for both engineering education and AI alignment research.
\end{abstract}


\section{Introduction}

Large Language Models (LLMs) are increasingly deployed across science, engineering, and education, where their ability to generate structured reasoning has shown rapid improvement \cite{openai2023gpt4, wei2022cot}. Yet in technical domains governed by symbolic rules and strict conventions, reliability depends not only on producing a numerically correct answer but on adhering to the methodological constraints that justify that answer. Unlike creative tasks, where plausibility may suffice, engineering disciplines demand \emph{procedure-faithful} reasoning: a solution derived using an incorrect sign convention or an invalid current direction is a failure even if the final number appears correct.

Most existing benchmarks for mathematical and scientific reasoning, such as GSM8K \cite{cobbe2021gsm8k}, MATH \cite{hendrycks2021math}, or ScienceQA \cite{lu2022scienceqa}, evaluate final-answer accuracy but rarely assess whether a model followed domain-specific rules. In electrical engineering, this omission is critical. Circuit analysis requires not only applying Kirchhoff's laws but also respecting arbitrary, user-imposed conventions (e.g., reference directions, mesh orientations)\cite{sadiku2021fundamentals, kuphaldt2010lessons, openstax2022college}. Violations of these conventions represent a distinct failure mode that final-answer evaluations cannot capture. We refer to this as \textbf{Convention Blindness}: the tendency of an LLM to override explicit instructions in favor of implicitly learned priors. This behavior creates a measurable dissociation between physical solving capability and instruction adherence. Recent survey work has similarly argued that contemporary AI benchmarks inadequately capture procedural and constraint-sensitive reasoning, emphasizing the need for evaluations that go beyond outcome correctness toward methodological fidelity \cite{ravishankara2025artificialintelligencecognitiveexamination}.

This phenomenon reflects a broader \textbf{neuro-symbolic reliability gap} observed in recent studies examining LLM failures in physics, control, and circuit reasoning \cite{skelic2025circuit, zhao2025mmcircuitevalcomprehensivemultimodalcircuitfocused, bubeck2023sparks}. While models increasingly master the computational aspects of physical laws, they often struggle when symbolic constraints conflict with patterns seen during pretraining. For example, a model may compute correct voltage drops but refuse to adopt a user-defined current direction if it contradicts statistically dominant patterns in its training data. In safety-critical applications—such as robotic actuation, embedded control, or power systems design—such sign errors can propagate into hazardous behavior despite superficially correct calculations.

To investigate this underexplored failure mode, we introduce \textbf{CircuChain}\footnote{To facilitate double-blind peer review, we provide an anonymized artifact bundle containing:
(i) the CircuChain problem set (50 base circuits with paired Control/Trap variants; 100 prompts total),
(ii) NGSPICE netlists, raw simulation outputs, and verification logs,
(iii) raw model responses used in our analysis, and
(iv) the scoring + judge pipeline (including the audit script).
Artifacts (pinned release): \href{https://anonymous.4open.science/r/CircuChain-IEEE-SoutheastCon-2026/}{https://anonymous.4open.science/r/CircuChain-IEEE-SoutheastCon-2026}}
, a diagnostic benchmark of 100 SPICE-verified circuit analysis tasks across five classical topologies. Unlike broad circuit datasets such as CIRCUIT \cite{skelic2025circuit} or MMCircuitEval \cite{zhao2025mmcircuitevalcomprehensivemultimodalcircuitfocused}, CircuChain is designed \emph{adversarially} to distinguish two orthogonal dimensions of LLM behavior: \emph{physical reasoning competence} and \emph{instruction compliance}. Each task includes a Control instance \cite{sadiku2021fundamentals, kuphaldt2010lessons, openstax2022college} and a counterbalanced \emph{Trap} instance, in which current directions, polarity assignments, or node labels are intentionally inverted. The goal is to reveal whether a model follows the prompt or defaults to training priors when the two conflict.

Evaluating five state-of-the-art models, including GPT-5 \cite{openai2025gpt5}, Claude Opus 4.5 \cite{anthropic2025opus45}, and o4-mini, we uncover a counter-intuitive \textbf{Compliance–Competence Trade-off}. Models with the strongest physical reasoning abilities demonstrate the highest rates of convention violations on Trap cases, whereas smaller models follow instructions more consistently but commit more fundamental physics errors. These results suggest that scaling alone does not resolve alignment challenges in mathematically rigid domains and highlight the need for new evaluation frameworks that test reasoning \emph{under constraint}.

Our contributions are threefold:
\begin{itemize}
\item \textbf{A Diagnostic Benchmark:} We introduce CircuChain, a controlled suite of 100 circuit-analysis tasks with paired Control/Trap variants designed to reveal reasoning–compliance discrepancies.
\item \textbf{A Multi-Stage Verification Pipeline:} We combine symbolic solving, SPICE simulation, and LLM-based error classification to distinguish physics errors from instruction-following errors.
\item \textbf{Empirical Identification of the Compliance–Competence Trade-off:} We show that stronger models may exhibit lower alignment with user-defined conventions, revealing a structural reliability gap in current LLMs.
\end{itemize}

\section{Related Work}
\subsection{LLMs in Scientific and Mathematical Reasoning}

Recent advances in Large Language Models have demonstrated impressive capabilities in mathematical and scientific reasoning, particularly when augmented with structured prompting techniques such as Chain-of-Thought (CoT) reasoning \cite{wei2022cot}. Models such as Minerva \cite{NEURIPS2022_18abbeef} and Galactica \cite{taylor2022galactica} show strong performance on symbolic derivations in mathematics and physics, while benchmarks like GSM8K \cite{cobbe2021gsm8k}, MATH \cite{hendrycks2021math}, and ScienceQA \cite{lu2022scienceqa} evaluate multi-step reasoning via final-answer correctness.

However, these evaluations primarily assess \emph{computational competence} rather than \emph{procedural fidelity}. In engineering domains, correctness is inseparable from the method used to derive an answer. A solution that violates sign conventions, reference directions, or methodological constraints remains invalid even if numerically correct. Our work extends prior reasoning benchmarks by explicitly testing whether models adhere to user-specified analytical procedures, revealing failure modes not captured by standard accuracy metrics.

\subsection{Alignment, Instruction Following, and Sycophancy}

A growing body of work has examined the alignment behavior of LLMs, particularly their tendency toward \emph{sycophancy}—agreeing with user assertions even when incorrect \cite{sharma2023sycophancy}. Instruction-following benchmarks and preference-modeling studies have shown that models may prioritize user satisfaction over factual correctness or safety constraints.

In contrast, our findings highlight an underexplored complementary failure mode: models that \emph{resist} user instructions when they conflict with dominant training priors. In rigid symbolic domains, such as circuit analysis, this manifests as the refusal to adopt arbitrary sign conventions or reference directions specified in the prompt. We refer to this phenomenon as \emph{Convention Blindness}. Rather than agreeing too readily with the user, highly capable models may instead privilege their internal representations of “typical” solutions, even when those representations contradict explicit instructions. CircuChain provides a controlled environment to study this inverse alignment failure in a physically grounded setting.

\subsection{Circuit and EDA Benchmarks}

Several recent benchmarks evaluate LLM performance on circuit understanding and electronic design automation (EDA) tasks. The \textit{CIRCUIT} benchmark \cite{skelic2025circuit} focuses on topology interpretation and conceptual reasoning across a diverse set of analog circuit problems. \textit{MMCircuitEval} \cite{zhao2025mmcircuitevalcomprehensivemultimodalcircuitfocused} and related datasets extend this paradigm to multimodal settings, incorporating schematic images, SPICE code generation, and design synthesis.

While these benchmarks are valuable for assessing broad competence, their multimodal nature conflates \emph{perceptual errors} (e.g., misreading diagrams) with \emph{reasoning errors}. In contrast, CircuChain intentionally employs text-only netlist descriptions to isolate symbolic reasoning and instruction compliance. By enforcing explicit analytical constraints—such as mesh current directions or polarity assignments—CircuChain evaluates a dimension of reliability largely unexplored by existing circuit benchmarks: whether models follow prescribed reasoning procedures when those procedures conflict with learned priors.

\section{Methodology}
CircuChain is designed to evaluate whether LLMs can follow explicitly stated circuit-analysis conventions while solving physically constrained problems. The benchmark emphasizes \emph{procedural fidelity}: adherence to user-defined sign conventions, variable definitions, and requested analysis method, in addition to numerical correctness.

\subsection{Dataset Structure and Topologies}
The benchmark comprises \textbf{5 fixed circuit topologies} and \textbf{50 manually authored problem instances} (10 per topology). Each instance is evaluated under two analysis modes—\textbf{Mesh/KVL} and \textbf{Nodal/KCL}—yielding \textbf{100 scored subtasks per model}. Each subtask requires reporting multiple variables (typically 3 mesh currents and 2 intermediate node voltages), enabling fine-grained failure attribution beyond final-answer accuracy.

The topologies were selected to span common complexity classes in linear DC circuit analysis:
\begin{itemize}
    \item \textbf{Topology 1: 3-Loop Supermesh.} Requires forming a supermesh and an auxiliary constraint equation induced by a shared source element.
    \item \textbf{Topology 2: 2-Mesh T-Network.} Tests shared-branch coupling and consistent application of the Passive Sign Convention (PSC).
    \item \textbf{Topology 3: Wheatstone Bridge.} A canonical bridge network testing node-voltage differences and balanced vs.\ unbalanced conditions.
    \item \textbf{Topology 4: Ladder Circuit.} A repetitive structure stressing variable consistency and coupled equation setup across stages.
    \item \textbf{Topology 5: Dependent Source Network (VCVS).} Includes a voltage-controlled voltage source, testing correct expression of dependency variables (e.g., $v_x$) in terms of primary state variables.
\end{itemize}

\subsection{Control vs.\ Trap Variants}
To probe failures that arise from reliance on stereotyped solution patterns (``Clever Hans'' behavior), each topology includes both \textbf{Control} instances (standard regimes) and \textbf{Trap} instances (counter-intuitive regimes). Trap variants preserve the \emph{same underlying topology template} but are constructed by altering parameter regimes and source orientations such that the physically correct solution violates common ``textbook'' intuitions (e.g., negative mesh currents under a clockwise definition, or reversed dominance across branches). In all cases, the required conventions (mesh direction, PSC, reference node, and variable names) are explicitly defined in the prompt.

Concretely, Trap instances are generated using one or more of the following mechanisms:
\begin{enumerate}
    \item \textbf{Source-dominance reversal:} Source magnitudes and/or polarities are chosen so that the net current flow opposes common left-to-right intuitions.
    \item \textbf{Asymmetric loading (stiff regimes):} Extreme resistance ratios (e.g., $R_1 \ll R_2$) create numerically stiff coupled equations that penalize informal simplifications.
    \item \textbf{Dependency stress (for Topology 5):} Dependent-source gains are set to induce strong coupling between controlling and controlled variables, making sign mistakes and omitted dependency terms immediately observable.
\end{enumerate}

\subsection{Prompting and Global Conventions}
All models are evaluated using a fixed prompt template to minimize prompt sensitivity. A global ``conventions'' block defines the default analysis rules (mesh direction, reference node, PSC) and every problem statement specifies the required method (Mesh/KVL or Nodal/KCL) and output variables. Models are instructed to return a structured \texttt{final\_answer} containing named numeric values for each requested variable to support automated grading.

\begin{figure}[htbp]
\centering
\fbox{\begin{minipage}{0.45\textwidth}
\footnotesize
\textbf{SYSTEM PROMPT (Excerpt):}\\
You are an expert Electrical Engineering Research Assistant. Solve the circuit problem step-by-step. Clearly define your Mesh/Nodal equations before solving.

\textbf{--- GLOBAL CIRCUIT CONVENTIONS ---}\\
1. \textbf{MESH CURRENTS:} Assume \textbf{CLOCKWISE} direction for all loops unless explicitly specified otherwise.\\
2. \textbf{REFERENCE NODE:} Measure nodal voltages with respect to the declared \textbf{GROUND (0V)} reference.\\
3. \textbf{PASSIVE SIGN CONVENTION:} Current entering the labeled (+) terminal of a passive element implies a positive voltage drop.\\
4. \textbf{DEPENDENT SOURCES:} Use the dependency definition stated in the problem; do not invent missing controlling variables.
\end{minipage}}
\caption{Excerpt of the fixed conventions block used across all evaluations.}
\label{fig:prompt_box}
\end{figure}

\subsection{Ground Truth Generation and Dual Verification}
For every problem instance, we generate a parameterized NGSPICE netlist by instantiating a fixed topology template and injecting the instance-specific component values and source orientations. We compute ground truth via DC operating point analysis (\texttt{.op}) and record node voltages and branch currents of interest. In parallel, we solve the same instance using an independent Python analytic solver (mesh and nodal equation solvers). We require agreement between NGSPICE and Python solutions within numerical tolerance for all recorded variables, yielding dual-verified ground truth for all dataset instances. No cases failed to converge.

\subsection{Automated Evaluation Pipeline}
We evaluate $5$ models on $100$ scored subtasks each (two analysis modes per instance), totaling $500$ scored subtasks.

\subsubsection{Stage 1: Solution Generation}
Models are queried via API at deterministic settings (temperature $=0$ or equivalent). Each prompt requires a structured \texttt{final\_answer} with named outputs for the requested variables.

\subsubsection{Stage 2: Numeric Scoring}
We parse the \texttt{final\_answer} values using a rule-based extractor; if parsing fails due to formatting, we use a constrained structured-extraction fallback. Each predicted variable is compared to ground truth using a hybrid tolerance:
\[
|x_{pred} - x_{truth}| \le \max(\epsilon_{abs}, \epsilon_{rel}|x_{truth}|),
\]
with small $\epsilon_{abs}$ to handle near-zero values. Subtasks are marked PASS only if all required variables satisfy tolerance.

\subsubsection{Stage 3: Failure Attribution (Judge Diagnosis)}
For failed subtasks, we use GPT-5 as an automated Judge for qualitative error attribution. To mitigate circular validation, we employ a \textbf{reference-guided} protocol: the Judge receives (i) the problem statement (including required conventions and method), (ii) the model's derivation, and (iii) the verified ground truth values, and is explicitly instructed \emph{not} to solve the circuit. The Judge assigns exactly one label:

\begin{itemize}
    \item \textbf{ERR\_SIGN\_CONVENTION:} Sign is incorrect despite correct magnitude, indicating violation of declared conventions (e.g., mesh direction or PSC).
    \item \textbf{ERR\_METHOD\_VIOLATION:} Used KVL when KCL was required, or vice versa.
    \item \textbf{ERR\_PHYSICS\_SETUP:} Incorrect KVL/KCL equation setup (fundamental modeling error).
    \item \textbf{ERR\_CALCULATION:} Correct equations but incorrect algebra/arithmetic.
    \item \textbf{ERR\_HALLUCINATION:} Invented components/values or produced non-responsive/unfinished work.
\end{itemize}

For aggregate reporting, we map these diagnoses into two high-level families:
\textbf{Compliance Errors} = \{\texttt{ERR\_SIGN\_CONVENTION}, \texttt{ERR\_METHOD\_VIOLATION}\} and
\textbf{Competence Errors} = \{\texttt{ERR\_PHYSICS\_SETUP}, \texttt{ERR\_CALCULATION}, \texttt{ERR\_HALLUCINATION}\}.


\section{Results}
We evaluate five models on CircuChain ($100$ scored subtasks per model; $50$ KVL and $50$ KCL). Each model response is graded against NGSPICE-verified ground truth, and failures are attributed using the judge rubric described in Section~III.

\subsection{Overall Performance}
Table~\ref{tab:leaderboard} reports overall accuracy. GPT-5 and Claude Opus 4.5 form a top tier with comparable performance (66\% vs.\ 65\%), followed by o4-mini (57\%). GPT-4o and GPT-4o Mini underperform on these multi-variable circuit tasks, suggesting a substantial gap between general-purpose chat models and reasoning-optimized variants.

To quantify uncertainty, we also report binomial 95\% confidence intervals (Table~\ref{tab:ci_overall}). At $N{=}100$, GPT-5 (66\% [56.3, 74.5]) and Claude Opus 4.5 (65\% [55.3, 73.6]) have strongly overlapping intervals, indicating no statistically meaningful separation at this sample size.

\begin{table}[htbp]
\caption{CircuChain Leaderboard (Judge-Labeled). Accuracy is the fraction of subtasks labeled \texttt{CORRECT}. $N{=}100$ per model (50 KVL, 50 KCL).}
\begin{center}
\begin{tabular}{lccc}
\toprule
\textbf{Model} & \textbf{Accuracy} & \textbf{KVL Acc.} & \textbf{KCL Acc.} \\
\midrule
GPT-4o Mini & 3.0\% & 4.0\% & 2.0\% \\
GPT-4o & 10.0\% & 20.0\% & 0.0\% \\
o4-mini & 57.0\% & \textbf{64.0\%} & 50.0\% \\
Claude Opus 4.5 & 65.0\% & 62.0\% & 68.0\% \\
GPT-5 & \textbf{66.0\%} & 60.0\% & \textbf{72.0\%} \\
\bottomrule
\end{tabular}
\label{tab:leaderboard}
\end{center}
\end{table}

\begin{table}[htbp]
\caption{Overall accuracy with binomial 95\% confidence intervals ($N{=}100$ per model).}
\begin{center}
\begin{tabular}{lcc}
\toprule
\textbf{Model} & \textbf{Acc.} & \textbf{95\% CI} \\
\midrule
GPT-5 & 66\% & [56.3, 74.5] \\
Claude Opus 4.5 & 65\% & [55.3, 73.6] \\
o4-mini & 57\% & [47.2, 66.3] \\
GPT-4o & 10\% & [5.5, 17.4] \\
GPT-4o Mini & 3\% & [1.0, 8.5] \\
\bottomrule
\end{tabular}
\label{tab:ci_overall}
\end{center}
\end{table}

\subsection{Robustness Under Trap Regimes (Control vs.\ Trap)}
To quantify robustness, we compare performance on Control vs.\ Trap regimes (Table~\ref{tab:reasoning_gap}). The dataset split is $n{=}52$ Control and $n{=}48$ Trap per model. GPT-4o shows the largest degradation under Traps, while o4-mini is relatively stable. Claude Opus 4.5 exhibits an \emph{inverted gap} (Trap $>$ Control), indicating strong robustness to the counter-intuitive parameter regimes used in our Trap construction.

\begin{table}[htbp]
\caption{Control vs.\ Trap Accuracy ($n{=}52$ Control, $n{=}48$ Trap per model). $\Delta =$ Control Acc. $-$ Trap Acc. (negative indicates Trap outperforms Control).}
\begin{center}
\begin{tabular}{lccc}
\toprule
\textbf{Model} & \textbf{Control Acc.} & \textbf{Trap Acc.} & \textbf{$\Delta$} \\
\midrule
GPT-4o Mini & 3.8\% & 2.1\% & +1.7\% \\
GPT-4o & 15.4\% & 4.2\% & +11.2\% \\
o4-mini & 57.7\% & 56.2\% & +1.4\% \\
Claude Opus 4.5 & 61.5\% & \textbf{68.8\%} & \textbf{-7.3\%} \\
GPT-5 & \textbf{69.2\%} & 62.5\% & +6.7\% \\
\bottomrule
\end{tabular}
\label{tab:reasoning_gap}
\end{center}
\end{table}

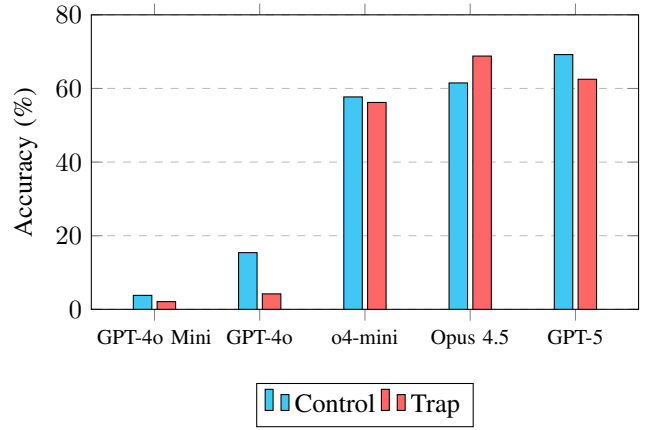
\begin{figure}[htbp]
\centering
\begin{tikzpicture}
    \begin{axis}[
        ybar,
        bar width=7pt,
        width=\linewidth,
        height=5.5cm,
        enlarge x limits=0.15,
        legend style={at={(0.5,-0.25)}, anchor=north, legend columns=-1},
        symbolic x coords={GPT-4o Mini, GPT-4o, o4-mini, Opus 4.5, GPT-5},
        xtick=data,
        ylabel={Accuracy (\%)},
        ymin=0, ymax=80,
        ymajorgrids=true,
        grid style=dashed,
        xticklabel style={font=\footnotesize, rotate=0},
    ]
    \addplot[fill=cyan!60] coordinates {(GPT-4o Mini, 3.8) (GPT-4o, 15.4) (o4-mini, 57.7) (Opus 4.5, 61.5) (GPT-5, 69.2)};
    \addplot[fill=red!60]  coordinates {(GPT-4o Mini, 2.1) (GPT-4o, 4.2)  (o4-mini, 56.2) (Opus 4.5, 68.8) (GPT-5, 62.5)};
    \legend{Control, Trap}
    \end{axis}
\end{tikzpicture}
\caption{Control vs.\ Trap accuracy on CircuChain ($n{=}52$ Control, $n{=}48$ Trap per model). Most models degrade under Trap regimes, while Claude Opus 4.5 exhibits an inverted gap (Trap $>$ Control).}
\label{fig:gap_chart}
\end{figure}

\subsection{Topology Sensitivity}
Aggregate accuracy can mask large differences across circuit classes. Table~\ref{tab:topology_acc} reports accuracy by topology. Each topology contributes $N{=}20$ subtasks per model (10 variations $\times$ two required methods). Performance is highly non-uniform: ladder circuits (Topology~4) are consistently solved by frontier models (GPT-5: 100\%, Opus: 95\%), whereas the 2-mesh opposing-source T-network (Topology~2) remains challenging even for top-tier models (GPT-5: 30\%, Opus: 50\%). This heterogeneity suggests topology-specific reliability gaps beyond what a single leaderboard score reveals.

\begin{table}[htbp]
\caption{Accuracy by topology (\%), $N{=}20$ subtasks per topology per model.}
\begin{center}
\begin{tabular}{lccccc}
\toprule
\textbf{Model} & \textbf{T1} & \textbf{T2} & \textbf{T3} & \textbf{T4} & \textbf{T5} \\
\midrule
GPT-5 & 85 & 30 & 50 & 100 & 65 \\
Claude Opus 4.5 & 60 & 50 & 45 & 95 & 75 \\
o4-mini & 60 & 20 & 65 & 80 & 60 \\
GPT-4o & 0 & 0 & 0 & 20 & 30 \\
GPT-4o Mini & 5 & 0 & 0 & 10 & 0 \\
\bottomrule
\end{tabular}
\label{tab:topology_acc}
\end{center}
\end{table}

\subsection{Compliance vs.\ Competence Failure Modes}
To connect performance to our central hypothesis (competence vs.\ compliance), we decompose failures using the judge taxonomy. We aggregate errors into:
\textbf{Compliance Errors} = \{\texttt{ERR\_SIGN\_CONVENTION}, \texttt{ERR\_METHOD\_VIOLATION}\} and
\textbf{Competence Errors} = \{\texttt{ERR\_PHYSICS\_SETUP}, \texttt{ERR\_CALCULATION}, \texttt{ERR\_HALLUCINATION}\}.
Table~\ref{tab:failure_agg} shows that GPT-5 failures are dominated by compliance (34\%), whereas Claude Opus 4.5 exhibits lower compliance error (7\%) but higher competence error (28\%). This supports a \emph{compliance--competence trade-off}: increasing raw problem-solving capability does not guarantee improved adherence to user-defined conventions.

\begin{table}[htbp]
\caption{Aggregated failure attribution (Judge-Labeled). Percentages are fractions of all subtasks ($N{=}100$ per model).}
\begin{center}
\begin{tabular}{lccc}
\toprule
\textbf{Model} & \textbf{Accuracy} & \textbf{Compliance Err.} & \textbf{Competence Err.} \\
\midrule
GPT-4o Mini & 3\% & 1\% & 96\% \\
GPT-4o & 10\% & 1\% & 89\% \\
o4-mini & 57\% & 31\% & 12\% \\
Claude Opus 4.5 & 65\% & 7\% & 28\% \\
GPT-5 & 66\% & \textbf{34\%} & \textbf{0\%} \\
\bottomrule
\end{tabular}
\label{tab:failure_agg}
\end{center}
\end{table}

\subsection{Judge Reliability Audit}
To validate the automated classification, we performed a secondary audit using Claude 3.5 Sonnet on a random sample of $N=50$ failure cases. The independent judge achieved a fine-grained agreement rate of \textbf{72.0\%} ($\kappa=0.53$) across the five specific error categories.

When aggregated into the primary \textbf{Compliance vs.\ Competence} dichotomy used for our main findings, agreement increased to \textbf{80.0\%} ($\kappa=0.57$). Most disagreements occurred at the boundaries of sub-categories (e.g., distinguishing a subtle Setup error from a Calculation error), which does not affect the primary trade-off conclusions.
\section{Discussion}

\subsection{The Compliance--Competence Trade-off}
CircuChain separates two orthogonal properties: (i) \emph{competence} in modeling and solving the circuit equations, and (ii) \emph{compliance} with user-specified conventions (e.g., mesh direction, reference polarity, and method constraints).
Across models, we observe a consistent divergence between these axes.
In particular, GPT-5 attains the highest overall accuracy (66\%) but exhibits a high rate of compliance failures (34\% of all subtasks), while showing no competence-attributed failures under our rubric (Table~\ref{tab:failure_agg}).
Conversely, Claude Opus 4.5 attains comparable accuracy (65\%) with substantially lower compliance error (7\%), but a higher rate of competence-attributed failures (28\%).

This pattern suggests that scaling raw problem-solving capability does not automatically improve adherence to arbitrary user constraints, a limitation increasingly attributed to the architectural decoupling of symbolic comprehension and execution \cite{zhang2025comprehension}.
A plausible interpretation is that models develop strong internal priors about ``standard'' solution conventions, and those priors can dominate when the prompt enforces less typical constraints.
This is consistent with broader findings that instruction-tuning and alignment methods improve helpfulness and task completion, but do not guarantee robust constraint satisfaction in edge cases or under competing objectives \cite{NEURIPS2022_b1efde53, bai2022constitutional}.
Rather than claiming a causal mechanism, we treat this as an empirical trade-off exposed by CircuChain: frontier capability and strict convention compliance can fail independently.

\subsection{Methodological Bias: Mesh vs.\ Nodal}
We also observe that models are not method-agnostic.
GPT-5 performs better on nodal (KCL) subtasks than mesh (KVL) subtasks (72\% vs.\ 60\%, Table~\ref{tab:leaderboard}), while GPT-4o fails all KCL subtasks (0\%).
One contributing factor is that nodal analysis often induces larger, fraction-heavy linear systems, amplifying algebraic sensitivity and increasing the probability of downstream arithmetic errors in token-based generation.
Another plausible factor is representational familiarity: SPICE-family simulators and modern circuit toolchains are grounded in (modified) nodal formulations, which may indirectly shape the distributions seen during model development \cite{ho1975modified}.
We present this as a \emph{methodological bias} hypothesis grounded in the observed performance asymmetry; establishing causality would require controlled ablations (e.g., isomorphic prompts rewritten to equalize algebraic form).

\subsection{Implications for Engineering Education and Safety-Critical Use}
The compliance failures we observe matter in real workflows because circuit analysis conventions are often \emph{assignment-scoped} and instructor-specific.
A tutoring system that returns a numerically plausible answer while silently violating the stated mesh direction or polarity convention can produce misleading feedback to learners, even when the learner is following the requested conventions.
For engineering education, this motivates tooling that (i) explicitly checks convention adherence, and (ii) surfaces convention mismatches as first-class errors rather than ``minor'' discrepancies.

For safety-critical engineering, the results motivate caution when LLM outputs are transformed into downstream artifacts (e.g., control logic, sign-sensitive code, or automated documentation).
In feedback and control settings, a sign or polarity inversion can change negative feedback into positive feedback and alter stability properties.
While CircuChain is not a control benchmark, the observed prevalence of sign-related failures in strong models highlights the need for explicit verification layers when deploying LLMs in sign-sensitive domains \cite{astrom2019feedback}.

\subsection{Limitations and Future Work}
CircuChain currently covers five linear circuit topologies and evaluates text-only reasoning under a fixed convention contract.
The task suite is intentionally diagnostic rather than exhaustive; expanding to richer topologies (e.g., op-amp circuits), AC/phasor analysis, and diagram-grounded inputs is a natural next step.
Second, failure attribution uses an LLM-based judge; while a secondary audit shows substantial agreement (fine-grained $\kappa \approx 0.53$; dichotomy $\kappa \approx 0.57$), future work should incorporate additional auditors and rule-based checks for key error modes.
Finally, extending CircuChain to compare performance under different convention contracts (e.g., explicit polarity flips, alternate reference nodes) would more directly stress convention alignment beyond parameter-based trap regimes.

\section{Conclusion}
We introduced \textbf{CircuChain}, a SPICE-verified benchmark designed to disentangle \emph{physical competence} from \emph{instruction compliance} in circuit analysis. Across five state-of-the-art models, we find that higher raw accuracy does not eliminate convention-sensitive failures. GPT-5 and Claude Opus 4.5 achieve comparable top-tier accuracy, yet their remaining errors differ qualitatively: under our judge-labeled aggregation, GPT-5 exhibits 0\% \emph{competence errors} but a high rate of \emph{compliance errors} (34\% of all subtasks), consistent with \textbf{Convention Blindness} under user-specified sign and method constraints. In contrast, Claude Opus 4.5 shows substantially lower compliance error (7\%) but higher competence error (28\%).

These findings reveal a reliability gap: numerical correctness is insufficient when user-declared conventions (e.g., mesh direction, polarity, required method) define validity. In safety- and specification-critical workflows (e.g., generating equations or control-oriented code from symbolic reasoning), sign or direction violations can invert intended behavior even when magnitudes appear correct. Future work should therefore move beyond accuracy-only metrics toward \textbf{compliance-aware evaluation and training}, rewarding solutions that are not only correct, but also verifiably consistent with declared engineering specifications.

\section*{Acknowledgment}
The authors used large language models to assist with the formatting and language improvement of this manuscript. All technical content was verified by the authors. The authors also gratefully acknowledge \textit{[Name Redacted for Double-Blind Review]} for providing infrastructure and financial support for this work.

\bibliographystyle{IEEEtran}
\bibliography{references}

\end{document}